\RequirePackage{amsmath}
\documentclass[12pt]{iopart}

\usepackage{graphics}
\usepackage{cite}
\usepackage{epstopdf}
\usepackage{ctable}
\usepackage{bm}
\usepackage{xcolor}
\usepackage{subfigure}
\newcommand\vect[1]{\bm{#1}}

\begin{document}
\title{Induced side-branching in smooth and faceted dendrites: theory and Phase-Field simulations}

\author{
Gilles Demange$^{1}$, Renaud Patte$^{1}$, Helena Zapolsky$^{1}$}

\address{$^{1}$GPM, CNRS-UMR 6634, University of Rouen Normandy, 76801 Saint \'Etienne Du Rouvray, France}

\vspace{2pc}

\ead{gilles.demange@univ-rouen.fr}

\begin{abstract}
The present work is devoted to the phenomenon of induced side branching stemming from the disruption of free dendrite growth. Therein, we postulate that the secondary branching instability can be triggered by the departure of the morphology of the dendrite from its steady state shape. Thence, the instability results from the thermodynamic trade-off between non monotonic variations of interface temperature, surface energy, kinetic anisotropy and interface velocity within the Gibbs Thomson equation. For purposes of illustration, the toy model of capillary anisotropy modulation is prospected both analytically and numerically by means of phase field simulations. It is evidenced that side branching can befall both smooth and faceted dendrites, at a normal angle from the front tip which is specific to the nature of the capillary anisotropy shift applied.
\end{abstract}


\maketitle

\section{Introduction}
\label{Intro}

Dendrites linger as a flagship example of patterning in nature and material science. These peculiar structures emerge during a myriad of phase and structural transformations, including the solidification of alloys and semiconductors \cite{arivanandhan2019crystallization}, the growth of ice crystals in undercooled water or vapor \cite{libbrecht2005physics,shimada2018rapid}, and the crystallization of lava at the boundary of the Earth's core \cite{PEPI,AG_ICB2013}.
Deciphering their growth process and ensuing morphology hitherto remains a stumbling block of non-linear physics, as well as a strong prerequisite to taylor the microstructural features determining the mechanical properties of industrial cast ingots \cite{hoyt2003}. In this regards, the secondary branching process receives a growing attention, insofar as side-branches set the length scales and patterns of microsegregation \cite{li1999evolution}, and contribute to the adjustment of the primary spacing during directional solidification \cite{hansen2002dendritic}. From a more putative perspective, the recurrent presence of secondary branched dendrites in the prebiotic soup might provide a plausible origin from primitive biological processes \cite{umantsev2011origin}. 

The fundamental understanding of dendrite growth relies on the pioneering works of Ivantsov \cite{ivantsov1952growth} which latches the P\'eclet number to the thermodynamical driving force, and the microscopic solvablity theory \cite{kessler1987growth,barbieri1989predictions,barbieri1989predictions} which provides a selection criterion  for the charactertistic wavelength of the tip \cite{Brener1990,brener1991pattern,alexandrov2016selection,alexandrov2013selection,alexandrov2017dendritic,kao2020stable,horvay1961dendritic,toropova2020non,JCG2021,PTA2021}. As for secondary branching, the conventional theory \cite{pieters1986noise,langer1987dendritic,barber1987dynamics,brener1995noise} postulates that side-branches stem from selective (thermal) noise amplification close to the tip. An alternative deterministic mechanism was proposed in \cite{glicksman2007deterministic}, which surmises that capillary or kinetic anisotropy induces time periodic non monotone temperature distribution along the tip. This causes velocity variations to befall the tip, thereby triggering side-branching. 


Now, blind spots remain concerning the description of side branching induced by the disruption of the dendrite growth process. And yet, dendrites grow under transient of perturbed growth conditions in practical casting processes  \cite{he2003heat} and in nature.  For instance, a smooth decrease of the velocity pulling on the directional solidification in Fe-C alloys was proved  to mitigate the coarsening of side branches in a PFM study \cite{xie2014growth}. Moreover, a strong finite and localized disturbance caused by a speck of dust was experimentally evidenced to cause the violent destabilization of the front tip, and the formation of abnormally fast growing side-branches \cite{gonzalez2004deterministic}. Finally, the successive decrement and increment of the supersaturation around the faceted tip of a snowflake branch entails the formation of induced side branching (ISB) \cite{libbrecht2019snow}, and multiplies the possible morphologies for ice crystals. The latter example in particular shades light on the possibility of induced side branching, in case of non local disturbance of the growth conditions of the propagating dendrite.

In this context, the present work addresses the phenomenon of induced side branching after non-local disturbance of the growth conditions. A simple mechanism for the secondary branching instability is proposed. It postulates that side-branching can be induced by an extended deviation of the dendrite shape from its steady state morphology ensuing growth conditions change. Therefrom, the destabilization of the tip is fueled  by a non-monotonic deviation of the interface temperature as prescribed by the Gibbs Thomson relation, which is compensated by localized modulations of the interface normal velocity. The studycase of capillary anisotropy modulation is selected as an analytically tractable toy model for a mismatched dendrite shape-growth condition couple. It is prospected for both smooth and faceted dendrite growth, before it is challenged by phase field simulations. As a result, branches are prone to form symmetrically  on both sides of the dendrite, at a given normal angle from the tip. Moreover, two distinct patterns are identified whether the capillary anisotropy parameter is increased or decreased at the transition on the one hand, and faceting steps in the process on the other hand.



\begin{figure}[ht]
\centering
\includegraphics[height=6cm]{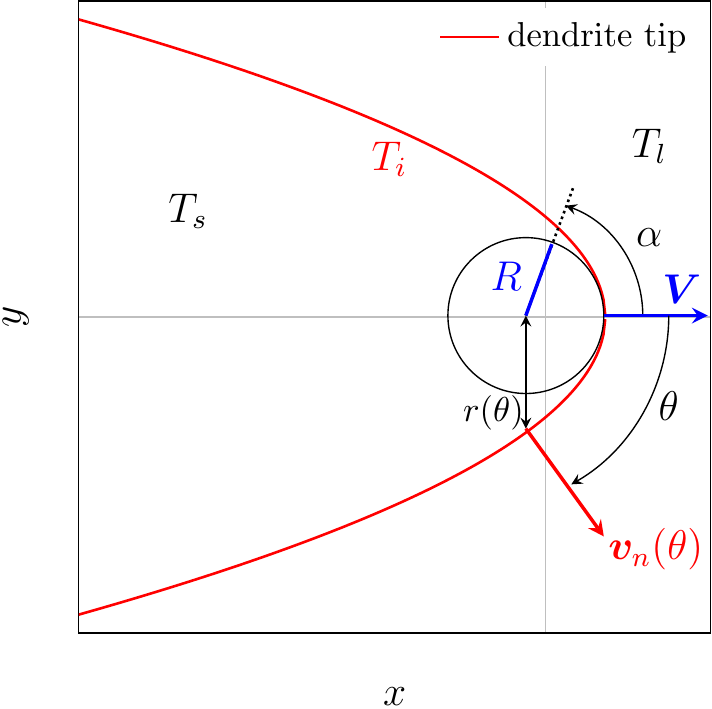}
\caption{Schematic representation of a dendrite tip propagating along $x$-axis, and having a parabolic dendrite shape with curvature radius $R$ and tip velocity $V$.}
\label{fig:tip} 
\end{figure} 	

\section{Theory: perturbation of Gibbs-Thomson relation}
\label{Theory}

\subsection{Governing equations}
\label{govern}

The present system of equations is based on the results obtained in \cite{alexandrov2018thermo}. For a dendrite growing in a pure undercooled liquid, the transport of heat in the stagnant melt follows the diffusion equation:

\begin{equation}
\label{eq1}
\frac{\partial T_l}{\partial t} = D_T\nabla^{(2)} T_l,
\end{equation}

\noindent
where $T_l$ is the temperature in liquid, and $D_T$ is the thermal diffusivity. Second, the heat condition is fulfilled at the solid/liquid interface:

\begin{equation}
\label{eq2}
T_Qv_n= D_T\left( \nabla T_s - \nabla T_l \right) \cdot \vect{n}.
\end{equation}

\noindent
where $T_s$ is the temperature in the solid,  and $T_Q = Q/c_p$ is the hypercooling (temperature for complete adiabatic solidification), $\vect{n}$ is the unit normal to the interface, and $v_n=\left( \vect{v}\cdot \vect{n} \right)$ will be referred as the interface normal velocity herefrom. Third, the Gibbs-Thomson condition holds at the solid/liquid boundary:

\begin{equation}
\label{eq3}
T_i=T_l=T_s=T_m - T_Qd(\vect{n})\mathcal{K}(\vect{n}) - T_Q\beta(\vect{n}) v_n ,
\end{equation}

\noindent
where $T_i$ is the phase transition temperature at the dendrite interface,  $T_m$ is the melting temperature for the pure system,  $\mathcal{K}(\vect{n})$ is the curvature of the interface, and $d(\vect{n})$ and $\beta(\vect{n})$ are the anisotropic capillary length and anisotropic kinetic coefficient, which root the anisotropic growth of the crystal. The far field condition is written in the form $T_l=T_{\infty}$ where $T_{\infty}$ is the fluid temperature far from the interface of the dendrite.

\subsection{Perturbation of the GT equation in two dimensions}
\label{GTperturb}

In two dimensions,  the anisotropic capillary length  $d(\theta)$ and the anisotropic kinetic coefficient $\beta(\theta)$ are specified for the case of $n$-fold lattice symmetry order:

\begin{equation}
\label{eq5}
\left\{
\begin{aligned}
d(\theta) &= d_0\left\{ A_d(\theta)+ A_d''(\theta)\right\}=d_0\left\{ 1-\alpha_d \cos\left[ n\left(\theta-\theta_d\right) \right] \right\}\\
\beta(\theta) &= \beta_0\left\{ A_\beta(\theta)+A_\beta''(\theta)\right\}= \beta_0 \left\{ 1-\alpha_\beta \cos\left[ n\left(\theta - \theta_\beta \right) \right] \right\},
\end{aligned}
\right.
\end{equation}

\noindent
where $d_0$ and $\beta_0$ are the capillary length and isotropic kinetic coefficient respectively, $A_{d,\beta}(\theta)=1+\alpha_{d,\beta}\cos\left[ n\left(\theta-\theta_{d,\beta}\right) \right]$ are the capillary and kinetic anisotropy functions, $\alpha_d$ and $\alpha_\beta$ are the small anisotropy parameters, and $\theta_d$ and $\theta_\beta$ designate the angles between the growth direction and the crystal orientations minimizing $d(\theta)$ and $\beta(\theta)$. The parameter $\alpha_d$ is tied to the anisotropy parameter $\epsilon_d$ by the relation $\alpha_d=(n^{2}-1)\epsilon_d$. 

The propagating 2D dendrite tip is sketched in figure \ref{fig:tip}, whereinto the interface normal velocity $v_n(\theta)$, the angle  $\theta=\arctan[n_y/n_x]$ between $x$-axis and the normal vector $\vect{n}$,  the polar angle $\alpha$, and the distance  between the center of the parabola's curvature circle $r(\theta)$ are indicated.

In the present work, the growth dynamics of the dendrite is disrupted by imposing a discontinuous shift of the growth conditions from the initial configuration (configuration 1), to a second configuration (configuration 2). All parameters and growth quantities are henceforth indexed accordingly (parabolic dendrite with tip radius $R^{(1,2)}$, curvature $\mathcal{K}^{(1,2)}$, anisotropy parameters $\alpha_d^{(1,2)}$ and $\alpha_\beta^{(1,2)}$ setting the anisotropic capillary length and anisotropic kinetic coefficient $d^{(1,2)}(\theta)$ and $\beta^{(1,2)}(\theta)$ respectively).  

This transition should result in a rebalancing of thermodynamical quantities stepping in GT condition (equation \ref{eq3}), including the interface normal velocity $v_n^{(1,2)}$ and tip velocity $V^{(1,2)}$, the curvature radius $R^{(1,2)}$, and the interface temperature $T_i^{(1,2)}$. For purposes of emphasizing the early perturbation of the transition on the interface velocity in the vicinity of the tip, the interface temperature will be assumed constant during the early stage of the transition. Upon equalizing GT equation \ref{eq3} for configuration 1 and 2, through the constant interface temperature $T_i$, the perturbed GT condition can be derived:

\begin{equation}
\label{eq10}
\delta v_n(\theta)=\frac{1}{T_Q\beta^{(2)}(\theta)}[T_m-T_i - T_Qd^{(2)}(\theta)\mathcal{K}^{(2)}(\theta)]-\frac{1}{T_Q\beta^{(1)}(\theta)}[T_m-T_i - T_Qd^{(1)}(\theta)\mathcal{K}^{(1)}(\theta)],
\end{equation}
where $\delta v_n(\theta)=v_n^{(2)}-v_n^{(1)}$.

\subsection{Simple analytical model for  side-branching induced by capillary anisotropy parameter shift in the case of the 6-fold symmetry}
\label{SBmodel}

In the present work, the perturbed GT equation \ref{eq10} is harnessed to elucidate how the swift change of one thermodynamical parameter may induce the disruption of the dendrite growth dynamics, one manifestation of which can be induced side branching (ISB).    

Among the thermodynamical parameters in equation \ref{eq10}, only the capillary anisotropy parameter was varied in the present study, insofar as it provides an analytically tractable expression for the resulting normal velocity shift  $\delta v_n$, as well as this parameter can be easily tuned in a numerical model, while keeping all other model parameters fixed. Some conclusions of this case should remain valid upon acting on other parameters, including kinetic anisotropy and tip radius. One step further, the specific studycase of the 6-fold symmetry for the dendrite was chosen in this work for numerical applications ($n=6$ in equation \ref{eq5}), so that $\alpha_d=35\epsilon_d$. In this context, the growth direction was aligned with  the $\vect{x}$-axis ($\theta_d=0$ and $\theta_\beta=\frac{\pi}{12}$  in equation \ref{eq5}). Finally, for the sake of simplicity, only the capillary contribution to the anisotropic growth was considered, for the kinetic anisotropy parameter $\alpha_\beta$ was set to zero.  Under such hypothesis, the following relations hold: $\beta^{(1)}(\theta)=\beta^{(2)}(\theta)=\beta_0$, and $R^{(1)}=R^{(2)}$ so that $\mathcal{K}^{(2)}(\theta)= \mathcal{K}^{(1)}(\theta)= \mathcal{K}(\theta)$. However, the capillary anisotropy constant varies: $\epsilon_d^{(1)}\neq \epsilon_d^{(2)}$, so that $d^{(1)}(\theta)\neq d^{(2)}(\theta)$. Accordingly, equation \ref{eq10} becomes:

\begin{equation}
\label{eq11}
\delta v_n(\theta)=\frac{\mathcal{K}(\theta)}{\beta_0}[d^{(1)}(\theta)-d^{(2)}(\theta)].
\end{equation}

\begin{figure}[ht]
\centering
\subfigure[~~smooth$\leftrightarrow$ smooth]{\includegraphics[width=4.2cm]{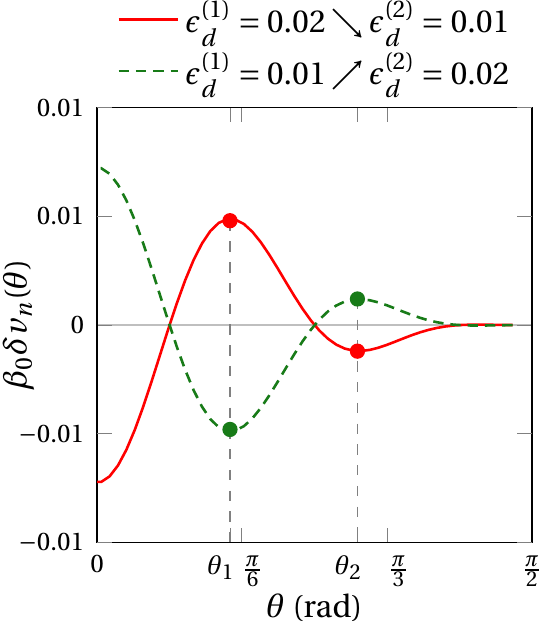}\label{fig:1smooth}}
\subfigure[~~smooth$\leftrightarrow$ facet]{\includegraphics[width=4.2cm]{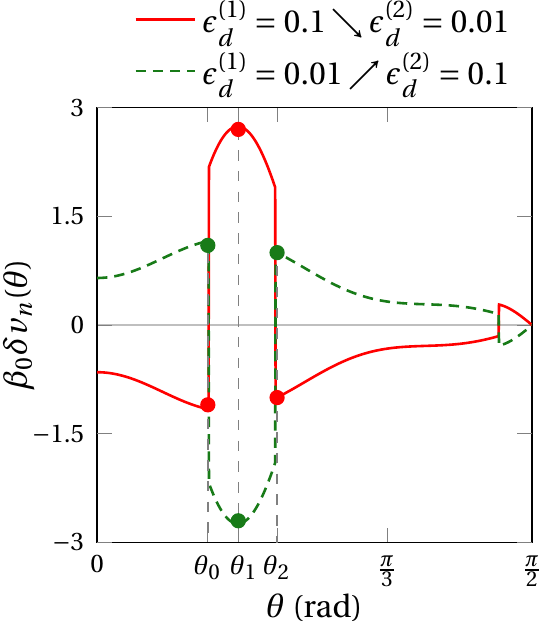}\label{fig:1facetsmooth}}
\subfigure[~~facet$\leftrightarrow$ facet]{\includegraphics[width=4.2cm]{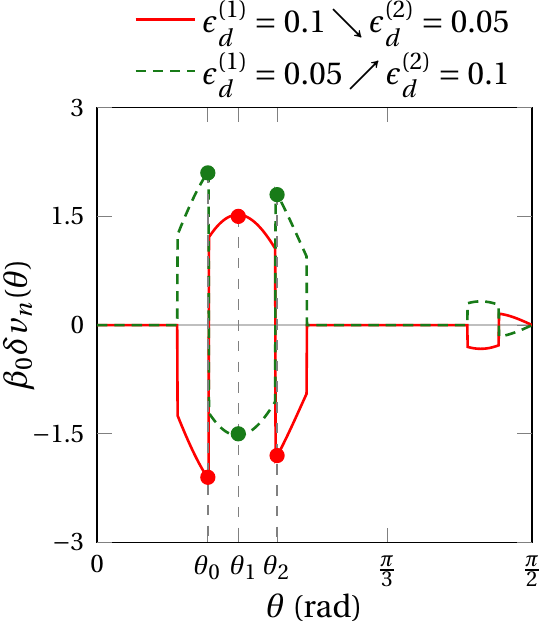}\label{fig:1facet}}
\caption{Variation of normal interface velocity $\delta v_n(\theta)$ as a function of the normal angle $\theta$ (in rad). Red line: decrease of capillary anisotropy parameter $\epsilon_d$. Green dashed line: increase of $\epsilon_d$. a) transition from smooth to smooth dendrite.  b) Transition from faceted (smooth) to smooth (faceted) dendrite. c)  Transition from faceted to faceted dendrite. Special angles $\theta_0$, $\theta_1$ ans $\theta_2$ correspond to local extrema of $\delta v_n$ where side-branching or carving is more likely.}
\label{fig:theo} 
\end{figure}

\paragraph{Smooth dendrite:} for a smooth (non-faceted) dendrite, the morphology of the tip can be approximated by Ivantsov's parabola of curvature radius $R$:

\begin{equation}
\label{eq6}
\frac{y(x)}{R}=1-\frac{1}{2}\left(\frac{x}{R}\right)^{2},
\end{equation}
whence assigning the curvature $\mathcal{K}$ the following analytical expression:

\begin{equation}
\label{eq7}
\mathcal{K}(x)\equiv \frac{y''(x)}{(1+(y'(x))^{2})^{3/2}}=\frac{1}{R\left(1+\left[\frac{x}{R}\right]^{2}\right)^{3/2}}=\frac{\cos(\theta)^3}{R}.
\end{equation}
Then, equation \ref{eq5} is used for the capillary distance $d(\theta)=d_0(A_d(\theta)+A_d''(\theta))$. Herein, equation \ref{eq11} simplifies to the following explicit expression for the interface normal velocity shift thereby induced: 

\begin{equation}
\label{C}
\delta v_n(\theta)=\frac{35d_0\cos^3(\theta)\cos\left(6\theta\right)}{\beta_0}\left[\epsilon_d^{(2)}-\epsilon_d^{(1)}\right]\\ 
\end{equation}
The corresponding profile of $\delta v_n(\theta)$ as a function of the normal angle $\theta$ is displayed in figure \ref{fig:1smooth}.

\paragraph{Faceted dendrite:} for a faceted dendrite,  equation \ref{eq5} should be amended. For (rough) faceted growth, a highly anisotropic surface energy ($\epsilon_d>1/35\equiv \epsilon_m$ for the 6-fold symmetry)  should be used. In this case, the surface energy anisotropy function $A(\theta)$ displays high energy directions in the interval $[-\theta_m,\theta_m]$ (modulo $\pi/3$ for the 6-fold symmetry), which generate unstable orientations in the equilibrium shape of the crystal (ECS).  These unstable orientations correspond to sharp corners in the ECS, associated with a negative stiffness $A_d+A_d''<0$ \cite{mcfadden1993phase}, and a negative capillary distance.

In this case, the mathematical problem of equations \ref{eq1}, \ref{eq2} and  \ref{eq3} becomes ill-posed, and $A(\theta)$ must be regularized.  In this work, the inverse polar plot of $A$ was convexified, following the procedure of Eggleston \cite{eggleston2001phase}. Unstable orientations can be determined by the tangent construction as per which the critical angle $\theta_m$ satisfies the following conditions: 

\begin{equation}
\label{thetam}
\frac{6\epsilon_d \sin(6\theta_m)}{\sin(\theta_m)}=\frac{1+\epsilon_d\cos(6\theta_m)}{ \cos(\theta_m)}.
\end{equation}
In the interval $[-\theta_m,\theta_m]$, the expression of the anisotropy function $A(\theta)$ becomes:

\begin{equation}
A_d(\theta)=A_1+B_1\cos(\theta), \quad \text{where:}\quad
\left\{
\begin{aligned}
&A_1=1+\epsilon_d \cos(6\theta_m)-B_1 \cos(\theta_m)\\
&B_1=\frac{6\epsilon_d \sin(6\theta_m)}{\sin(\theta_m)},
\end{aligned}
\right.
\label{Aregul}
\end{equation}
thereby ensuring the continuity of $A_d(\theta)$ and its first derivative $A_d'(\theta)$, and a positive value of the capillary distance $d_0(A_d(\theta)+A_d''(\theta))\geq 0$ defined piece-wise thereafter by:

\begin{equation}
\label{eq5facet}
d(\theta) = 
\left\{
\begin{aligned}
&d_0\left\{ 1-\alpha_d \cos\left[ n\left(\theta-\theta_d\right) \right] \right\} \quad \text{for}  \quad \theta \in [-\theta_m,\theta_m]\\
&1+\epsilon_d \cos(6\theta_m)-\frac{6\epsilon_d \sin(6\theta_m)}{\sin(\theta_m)}\cos(\theta_m) \quad \text{otherwise}.
\end{aligned}
\right.
\end{equation}
The corresponding profiles of $\delta v_n(\theta)$ provided by equation \ref{eq11} are displayed in figures \ref{fig:1facetsmooth} and \ref{fig:1facet} for smooth$\leftrightarrow$ facet and facet$\leftrightarrow$ facet transitions respectively, under the approximation $\mathcal{K}(\theta)\simeq \cos(\theta)^3/R_{\text{moy}}$, where $R_{\text{moy}}$ is an effective curvature radius for the faceted tip. The implications of this simplification will be discussed hereafter.

In figure \ref{fig:1smooth}, the decrease of the anisotropy parameter $\epsilon_d$ (red curve, $\epsilon_d^{(1)}=0.02$, $\epsilon_d^{(2)}=0.01$) leads to the shrinking of tip velocity $V=v_n(\theta=0)$, the rise of the normal velocity $v_n(\theta)$ in the vicinity of angle $\theta_1\simeq 0.48$ for the present setting of $\epsilon_d^{(1)}$ and $\epsilon_d^{(2)}$ (red dot,  $\delta v_n(\theta_1)>0$), and the small deceleration of the interface close to the angle $\theta_2\simeq 0.94$ (red dot, $\delta v_n(\theta_2)<0$). 

On the contrary, when  anisotropy parameter $\epsilon_d$ is incremented (green dashed curve, $\epsilon_d^{(1)}=0.01$, $\epsilon_d^{(2)}=0.02$), the situation is inversed, as $\delta v_n(\theta_1)<0$ and $\delta v_n(\theta_2)>0$. In this work, we surmise that the anisotropic variation of the normal velocity envisioned in figure \ref{fig:1smooth}, which stems from the step variation of the capillary anisotropy parameter, might as well enable the induced secondary branching of the dendrite. 

In addition, we suggest that hiking or lessening  the anisotropy parameter $\epsilon_d$ leads to two distinct side branching configurations. Whenever $\epsilon_d$ is decremented, the acceleration of the interface normal displacement is highest ($\delta v_n(\theta_1)>0$) at the normal angle $\theta_1$, versus $\theta_2\simeq 2\theta_1$ when $\epsilon_d$ is decreased. As a consequence, the initially induced secondary branching sprout is formed closer to the tip in the first case, before it propagates outward from the tip. Now, in the so-called linear region spanning a few tip radius from the dendrite tip \cite{langer1987dendritic}, the amplitude of the secondary branch grows exponentially from the tip \cite{gonzalez2004deterministic}, so that the closer to the tip the side brand is generated, the more it will thrive hereafter, and the stronger the ISB. 

For this reason, the decrease of $\epsilon_d$ is less likely to induce secondary branching. However, the decrease of $\epsilon_d$ also causes the strong deceleration of interface normal displacement in the normal direction  $\theta=\theta_1$, resulting in a hollow region that precedes the $\theta$ region favorable to the formation of a side branch. We suggest that this hollow portion might eventually curve undercooling isolines toward a slightly stronger undercooling in the $\theta_2$, whence fostering side-branching. 

A similar behavior can be expected for the  smooth$\leftrightarrow$ facet dendrite transition (figure \ref{fig:1facetsmooth}), notwithstanding two significant differences. First, the second extremum $\theta_2=\pi/3-\theta_m\simeq 0.65$ for the present setting of $\epsilon_d^{(1)}$ and $\epsilon_d^{(2)}$ corresponds to the lower bond of unstable orientations centered in $\pi/3$. In this regards, it is located closer to the dendrite tip than for the smooth$\leftrightarrow$ smooth transition ($\theta_2\simeq 0.94$). As a consequence, it is more likely to result in side-branching when $\epsilon_d$ is increased (faceted $\to$ smooth dendrite). Second, an additional extremum in  $\theta_0=\theta_m\simeq 0.4$ corresponding to the higher bound of the unstable orientations in the vicinity of the $\theta$ direction is located, which might hinder the secondary branching process when $\epsilon_d$ is decreased.

The latter observations remain true for the facet$\leftrightarrow$ facet dendrite transition (figure \ref{fig:1facet}).

\section{Phase-Field simulations}
\label{Simulation}

\subsection{Phase-Field model}
\label{PFM}

In this work, we simplified the solidification model introduced in \cite{demange2017phase} for strongly anisotropic dendrite growth in three dimensions, to the  two-dimensional case:

\begin{align}
A(\theta)^2\partial_t \psi&= -f'(\psi) - \lambda g'(\psi)\Theta \label{1}\\
&+\frac{1}{2}\nabla \cdot\left(|\nabla\psi|^2 \frac{\partial \left[A(\theta)^2\right]}{\partial \nabla\psi} + A(\theta)^2\nabla \psi\right),\nonumber\\
\partial_t \Theta &=\widetilde{D}\nabla^2 \Theta+\frac{L}{2c_p} \partial_t \psi.\label{2}
\end{align}
Here, $\psi$  is the order parameter ($=1$ for the solid phase, and -1 for the liquid), and $\Theta=(T-T_m)/T_Q$ is the reduced undercooling. In equation (\ref{1}), the double well potential $f(\psi)=-\psi^2/2+\psi^4/4$ is the free energy density of the system, at melting temperature $T_m$, and $g'(\psi)=(1-\psi^2)^2$ is an interpolation function. The model is equipped with an anisotropic function $A(\theta)=1+\epsilon_d\cos(6\theta)$ for smooth dendrite growth, the normal angle $\theta$ can be calculated from  the unit normal vector $\vect{n}=-\nabla \psi/|\nabla \psi|$ of $\psi$, as $\theta=\arctan[n_y/n_x]$.

\begin{figure}[ht]
\centering
\subfigure[~~WS]{\includegraphics[width=4cm]{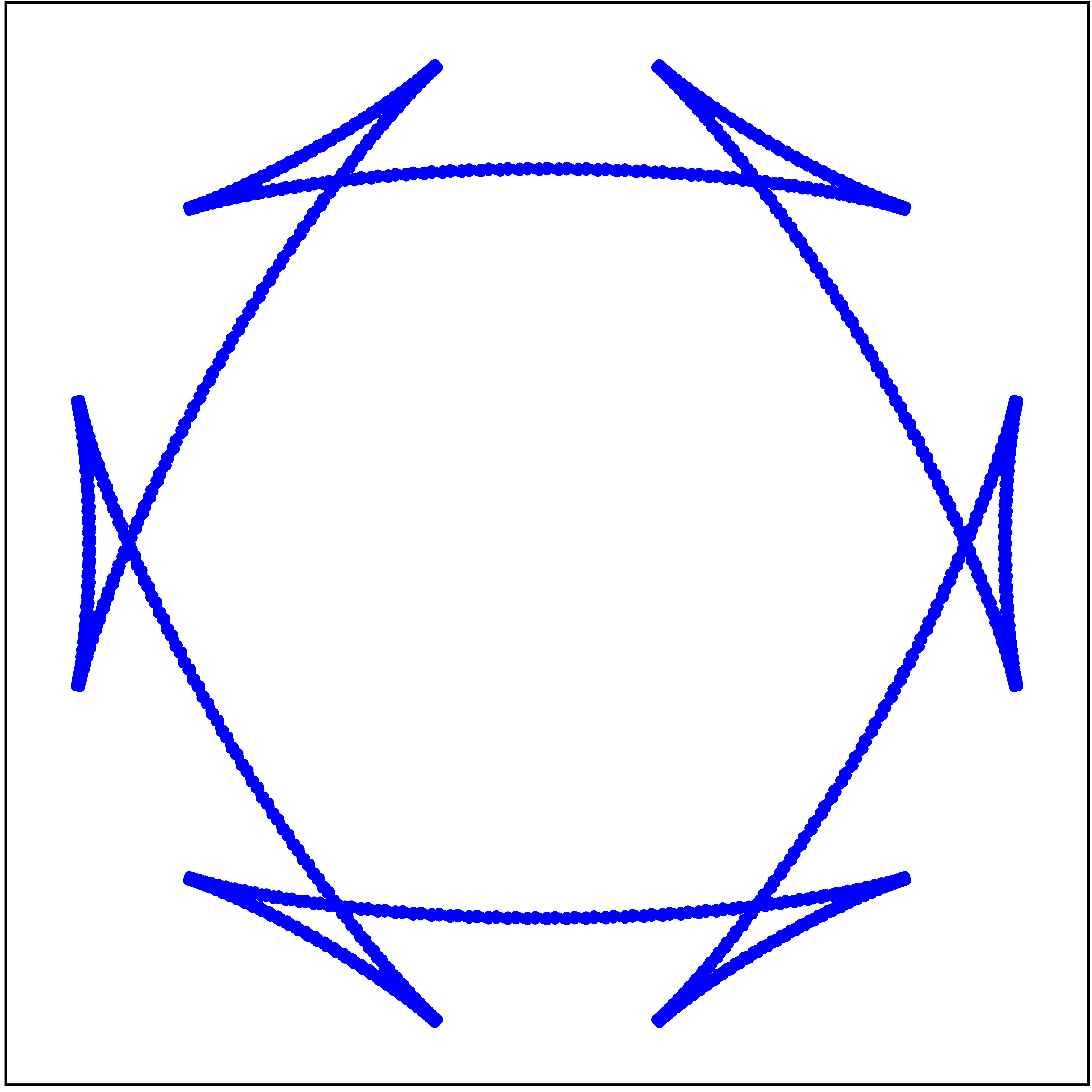}\label{fig:wulffa}}\hspace{0.2cm}
\subfigure[~~ECS]{\includegraphics[width=4cm]{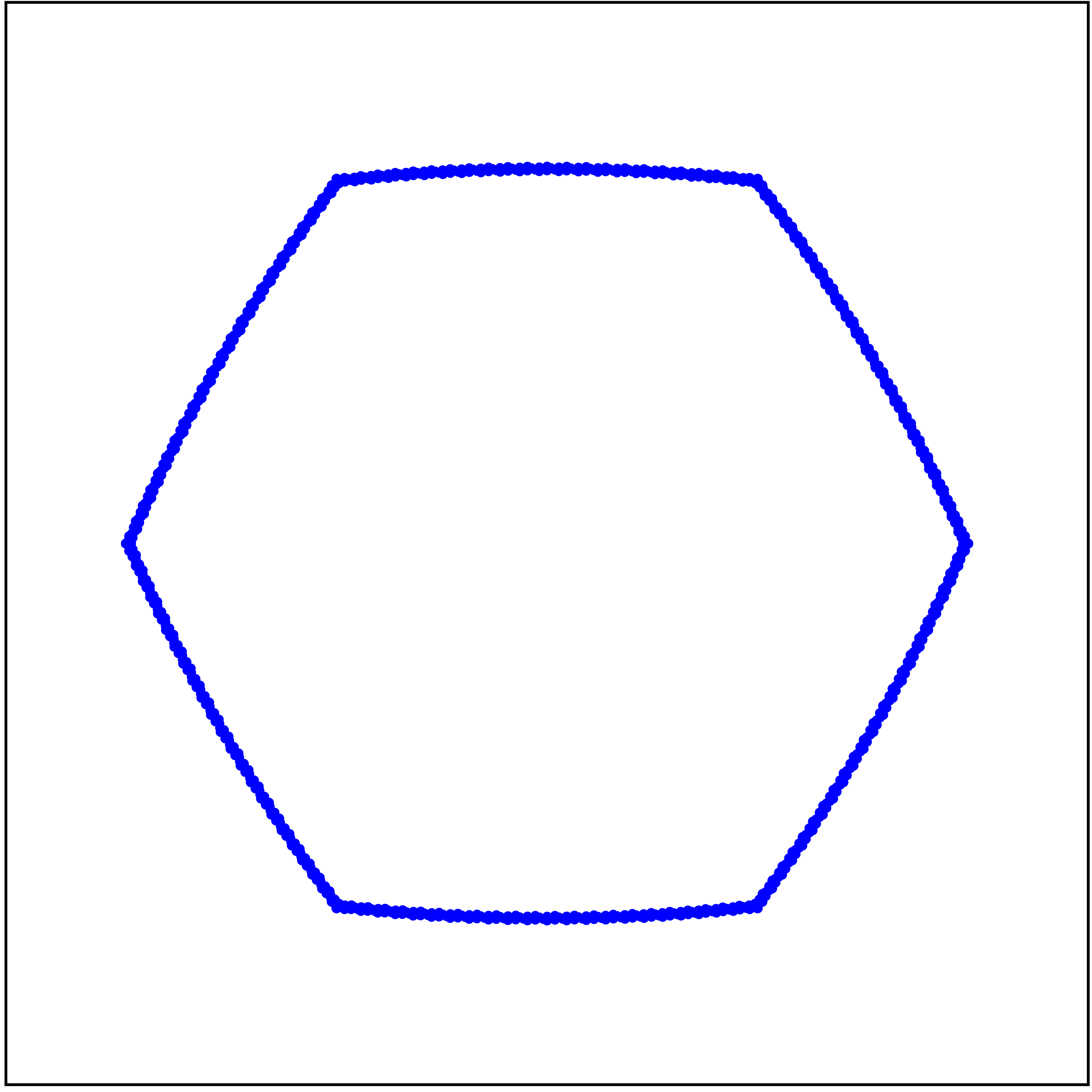}\label{fig:ecsa}}\hspace{0.2cm}
\subfigure[~~$1/A$]{\includegraphics[width=4cm]{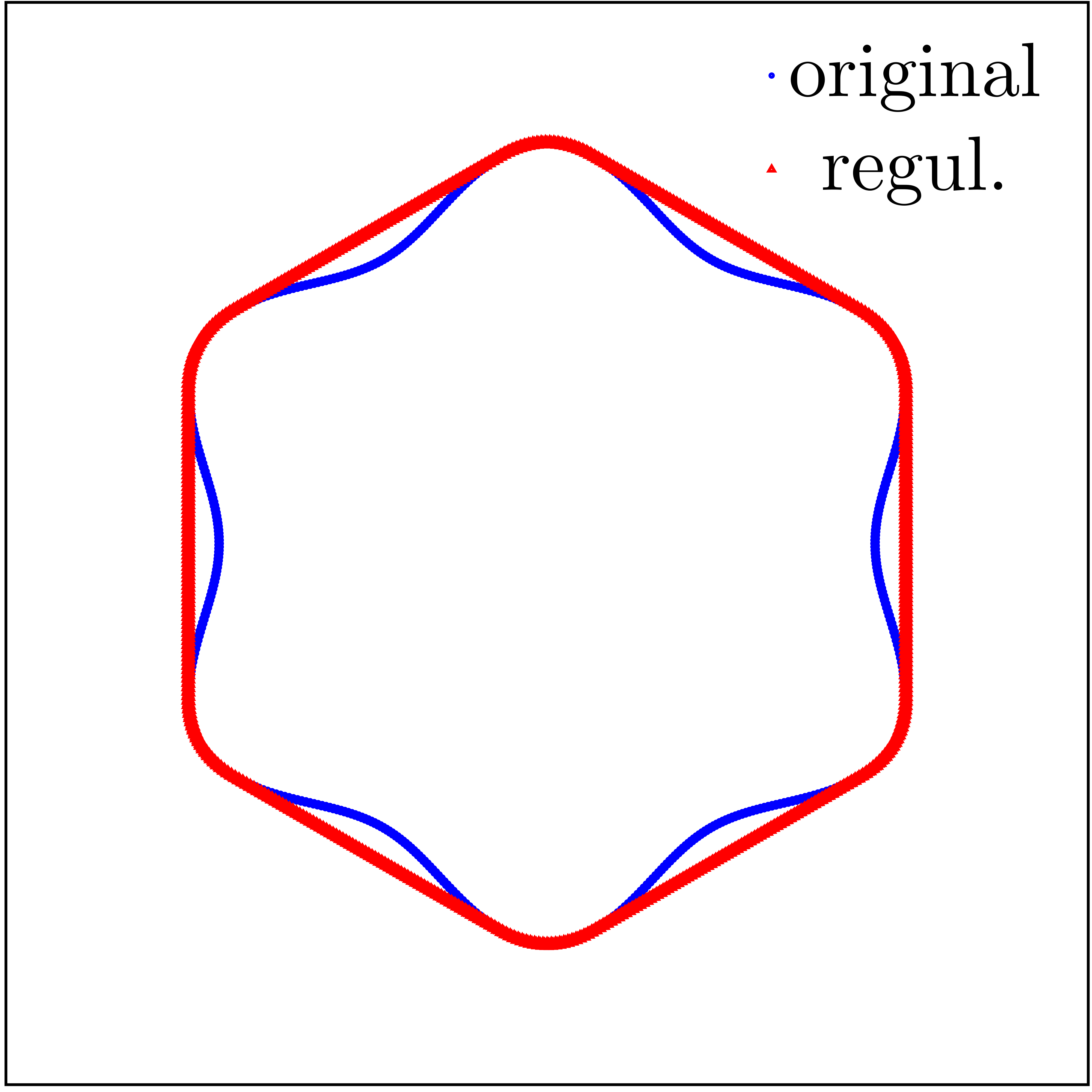}\label{fig:inva}}
\caption{Regularization procedure for $\epsilon_d=0.1$. a) Wulff shape (WS) before regularization. b) Equilibrium crystal shape (ECS) after regularization. c) Original and confexified inverse polar plot of $A_d$. Figres are taken from a previous publication  \cite{demange2017growth}.}
\label{fig:steps} 
\end{figure}

For faceted growth, the regularization algorithm introduced in the previous section  was used. The result of the regularization procedure is summarized in figure\ref{fig:steps} for $\epsilon_{d}=0.1>\epsilon_m$. Before regularization, the 2D Wulff shape (WS) for the ECS displays ears corresponding to unstable orientations (figure \ref{fig:wulffa}). The associated inverse 2D polar plot is concave in these directions (red curve in figure  \ref{fig:inva}).  After regularization, the ears are removed (figure \ref{fig:ecsa}), and the inverse polar plot becomes convex (blue curve in figure  \ref{fig:inva}). Besides that, the resulting 2D polar plot is regular and coincides with the original 2D polar plot where the curvature was already positive before the regularization procedure (red curve in figure  \ref{fig:inva}).

Equations \ref{1} and \ref{2} were solved in reduced units using periodic boundary conditions. Simulations were performed using the grid spacing $\Delta x=0.8$, and the time step $\Delta t=0.05$, in the bosom of a 3000 $\times$ 800 simulation box. This choice guarantees that the spacing between the dendrite tip and the boundary is several times the diffusion length during the simulation. The initial condition was a spherical germ of radius $4\Delta x$.  The coupling constant was set to $\lambda=3.0$  \cite{ramirez2004phase}. Simulations displayed in this work were performed at reduced undercooling $\Theta_{\infty}=0.65$, corresponding to the diffusion limited growth regime.
All results are given in reduced space and time units.

\subsection{PFM simulations: the studycase of capillary effects perturbation}
\label{PFMsimu}

\begin{figure}[ht]
\centering
\subfigure[$\epsilon_d^{(1)}=0.02\searrow \epsilon_d^{(2)}=0.01$]{\includegraphics[width=14cm]{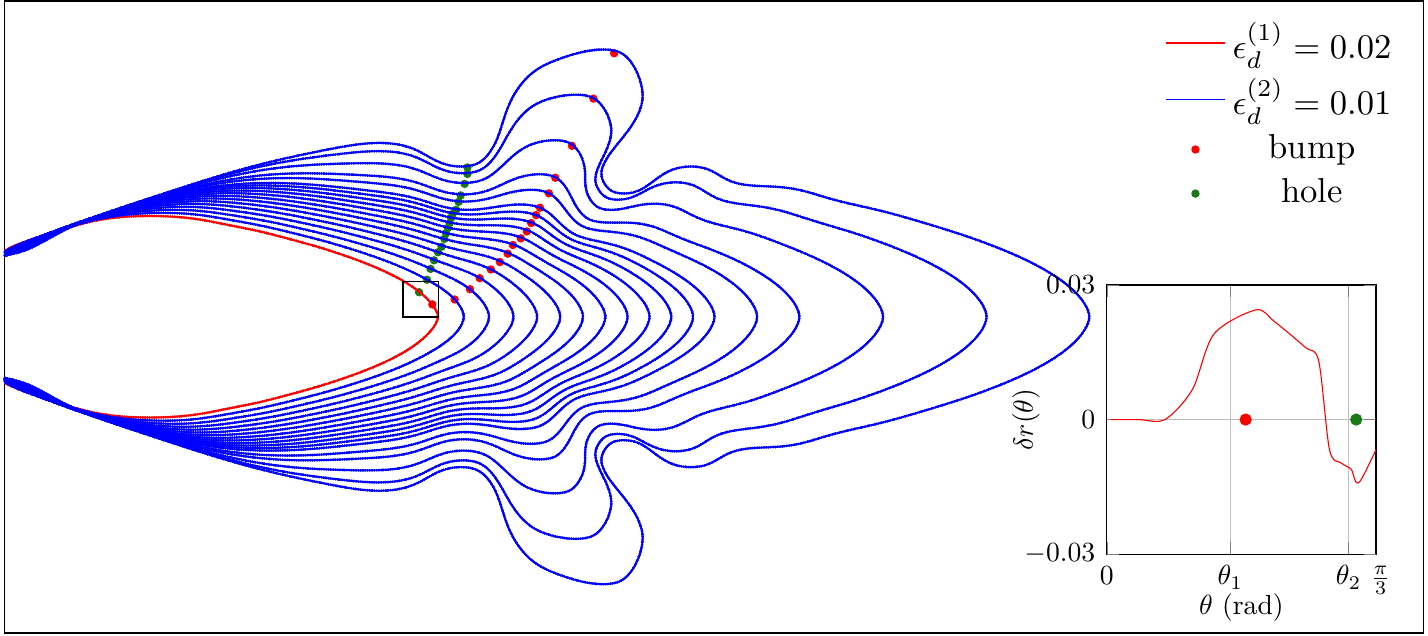}\label{fig:C}}
\subfigure[$\epsilon_d^{(1)}=0.01\nearrow \epsilon_d^{(2)}=0.02$]{\includegraphics[width=14cm]{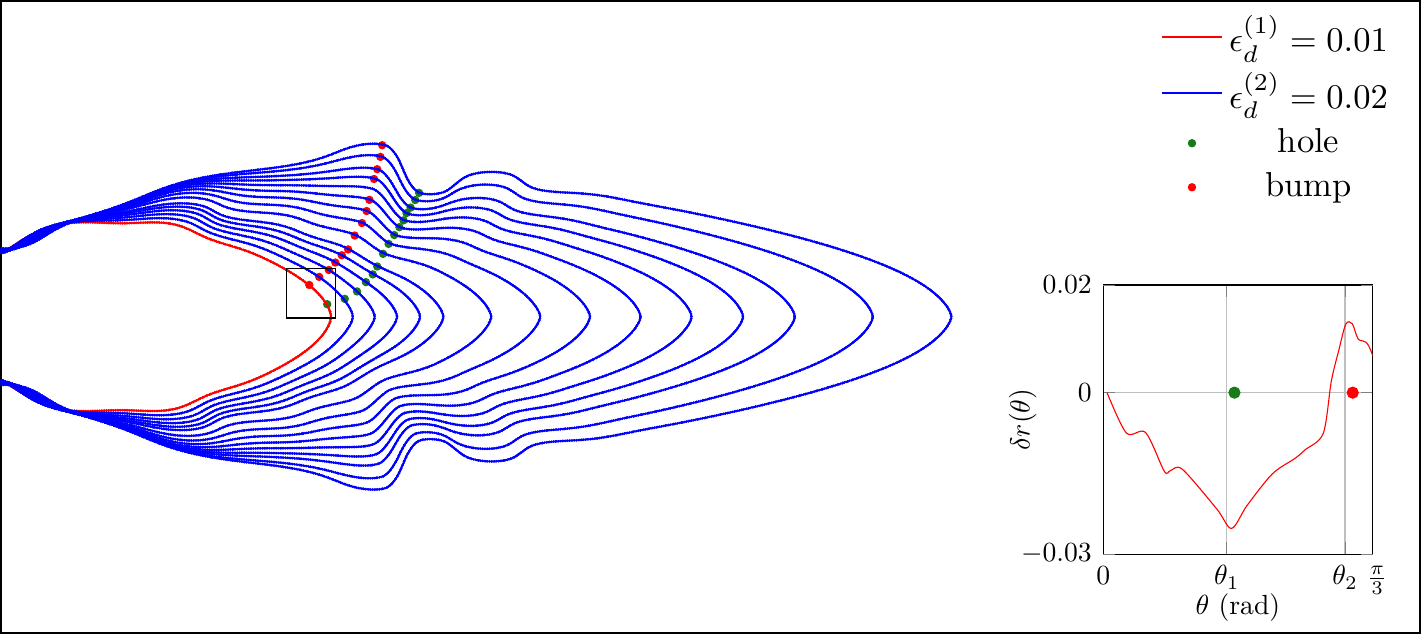}\label{fig:D}}
\caption{Dendrite morphological evolution from smooth tip to smooth tip after discontinuous change of anisotropy parameter $\epsilon_d$ at reduced time $t=400$. a) From $\epsilon_d^{(1)}=0.02$ corresponding to a smooth parabolic dendrite with steady state tip radius $R=9.6$ (red curve, $t=400$) to $\epsilon_d^{(2)}=0.01$ corresponding to a smooth parabolic dendrite with steady state tip radius $R=13.2$ (blue curves from $t=400$ to $t=1000$).  b) From $\epsilon_d^{(1)}=0.01$ (red curve, $t=400$, $R=13.2$) to $\epsilon_d^{(2)}=0.01$  (blue curves from $t=400$ to $t=1000$, $R=9.6$). a-b) Insert: initial variation of radius $r(\theta)$ between $t=400$ and $t=402.5$.}
\label{fig:smooth} 
\end{figure} 	

\begin{figure}[ht]
\centering
\subfigure[$\epsilon_d^{(1)}=0.1\searrow \epsilon_d^{(2)}=0.01$]{\includegraphics[width=14cm]{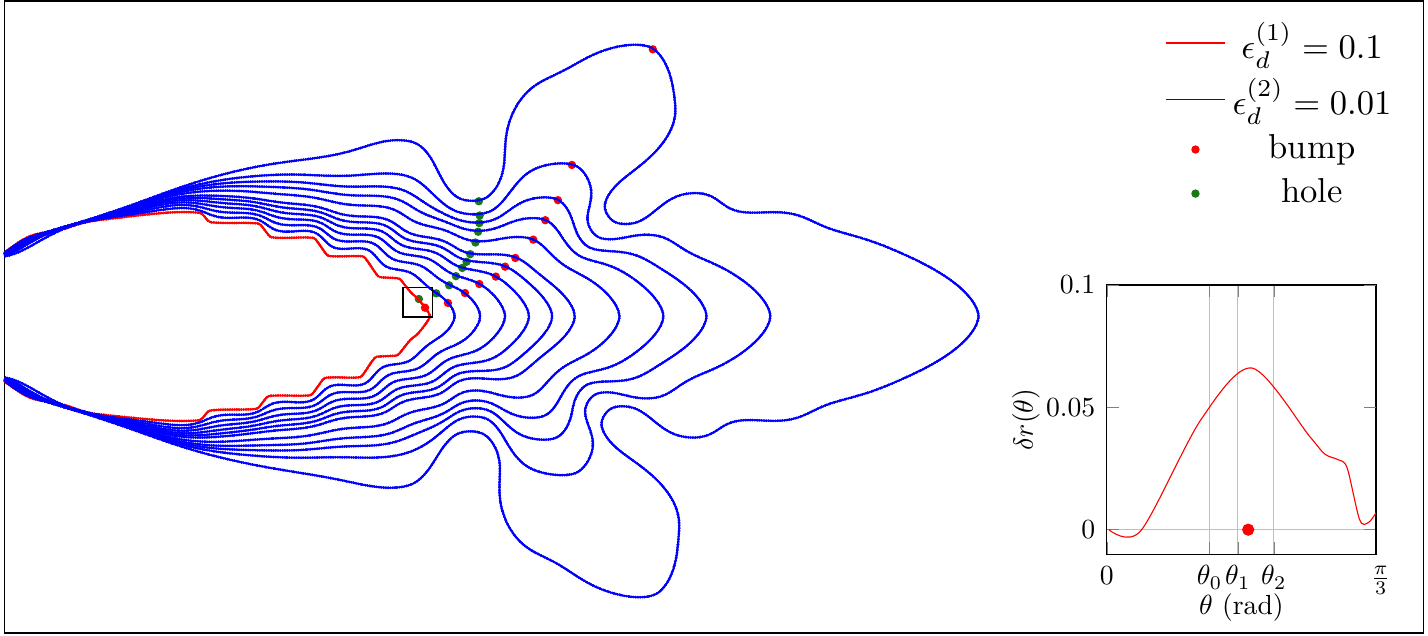}\label{fig:A}}
\subfigure[$\epsilon_d^{(1)}=0.01\nearrow \epsilon_d^{(2)}=0.1$]{\includegraphics[width=14cm]{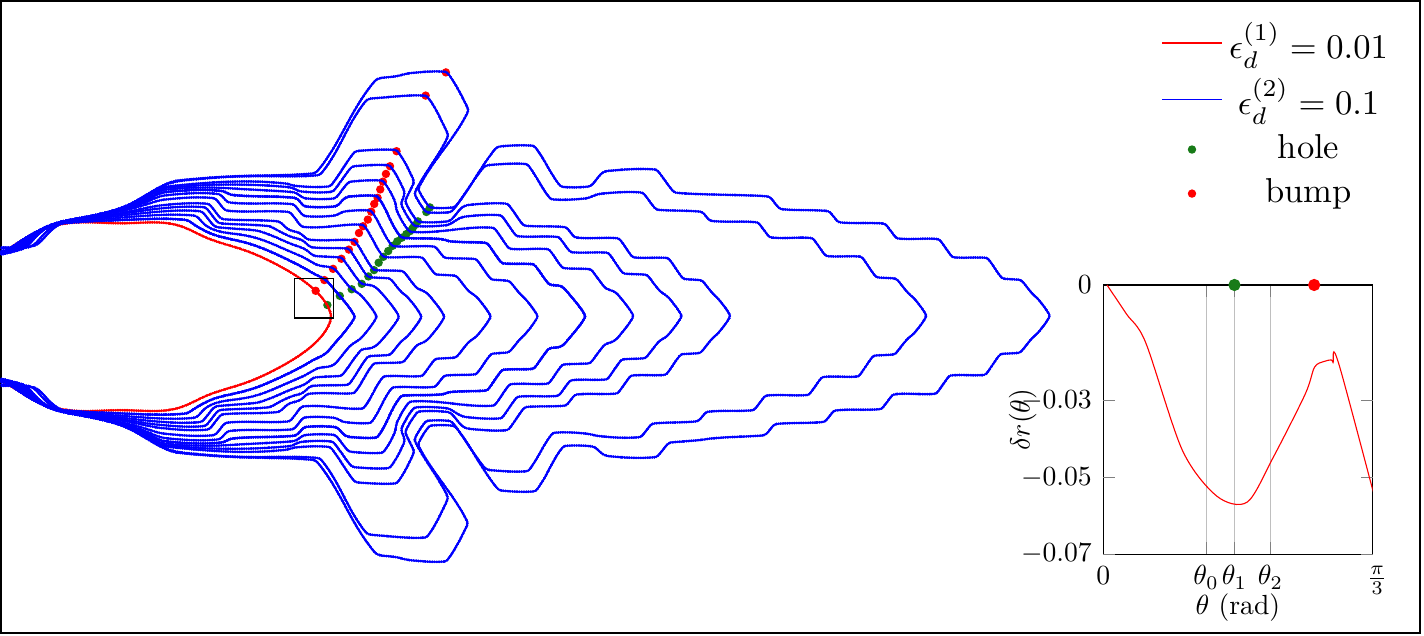}\label{fig:B}}
\caption{Dendrite morphological evolution from faceted (smooth) tip to smooth (faceted) tip after discontinuous change of anisotropy parameter $\epsilon_d$ at reduced time $t=400$. a) Decrease of $\epsilon_d$  from $\epsilon_d^{(1)}=0.1$ corresponding to a faceted dendrite with steady state tip radius $R=2.9$ (red curve, $t=400$) to $\epsilon_d^{(2)}=0.01$  (blue curves from $t=400$ to $t=1000$, $R=13.2$). b) Increase of $\epsilon_d$  from $\epsilon_d^{(1)}=0.01$ (red curve, $t=400$, $R=13.2$) to $\epsilon_d^{(2)}=0.1$  (blue curves from $t=400$ to $t=1000$, faceted dendrite with $R=2.9$). a-b) Insert: initial variation of radius $r(\theta)$ between $t=400$ and $t=402.5$.}
\label{fig:facetsmooth} 
\end{figure} 	

\begin{figure}[ht]
\centering
\subfigure[$\epsilon_d^{(1)}=0.1\searrow \epsilon_d^{(2)}=0.05$]{\includegraphics[width=14cm]{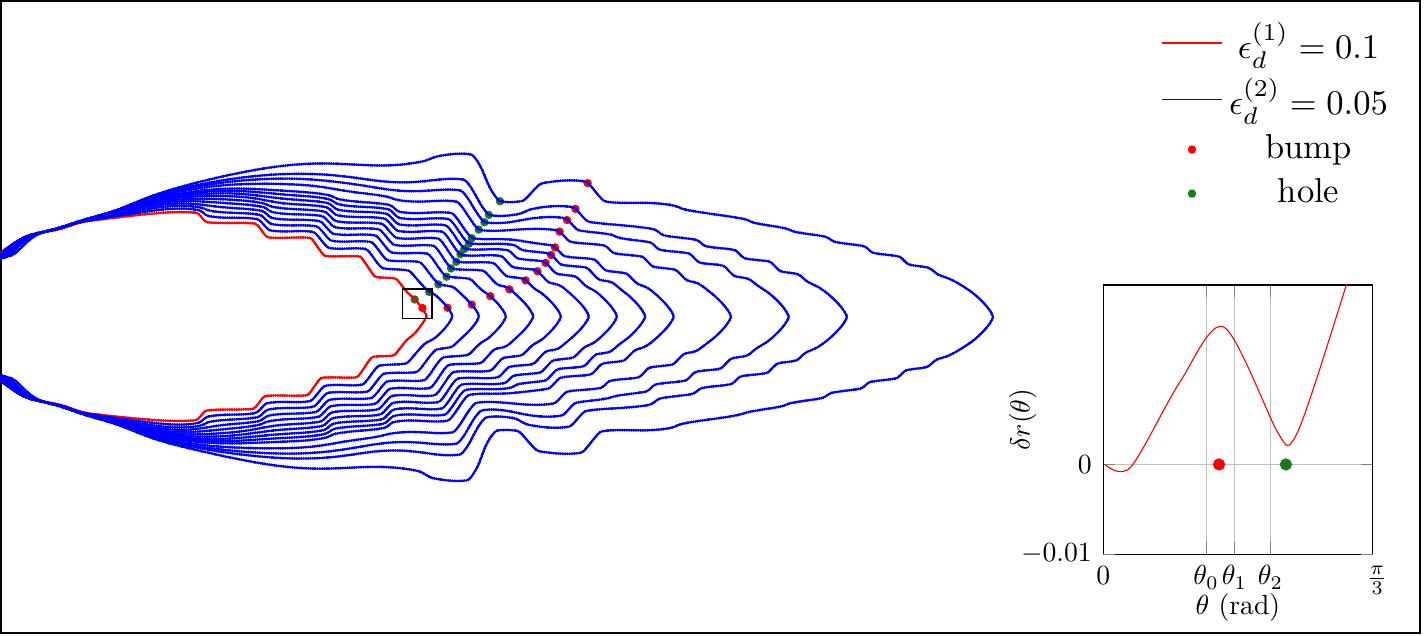}\label{fig:E}}
\subfigure[$\epsilon_d^{(1)}=0.05\nearrow \epsilon_d^{(2)}=0.1$]{\includegraphics[width=14cm]{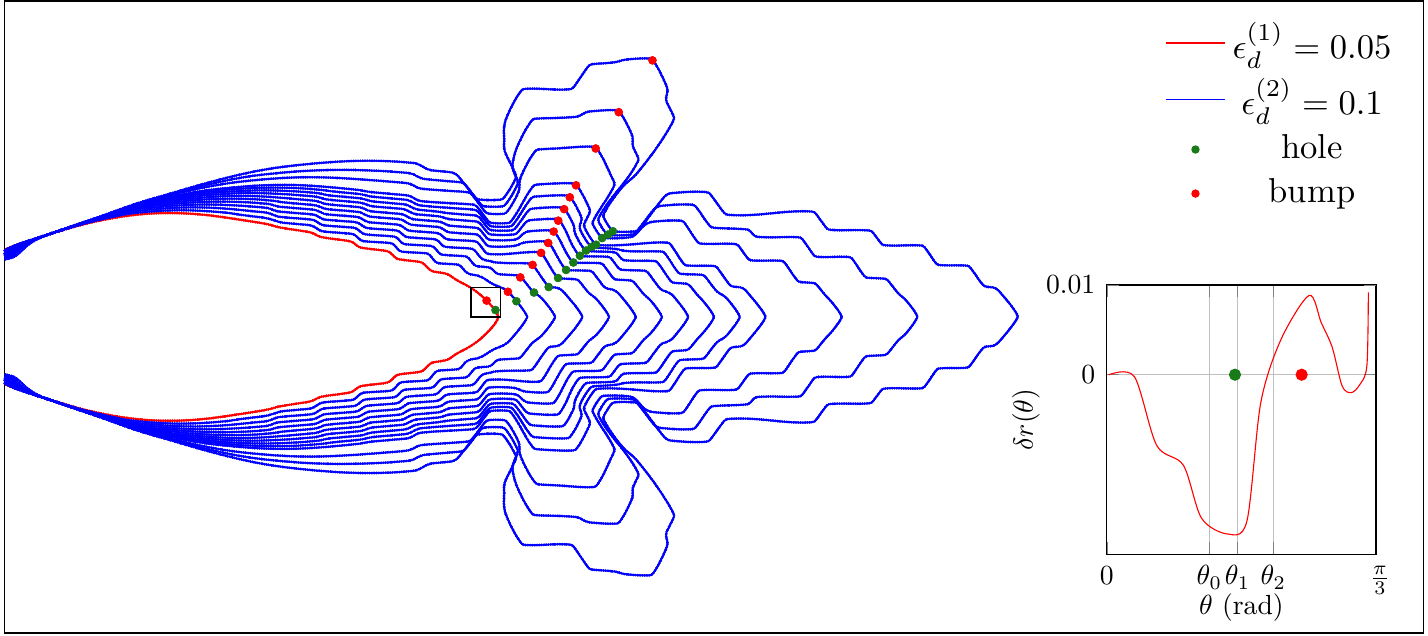}\label{fig:F}}
\caption{Dendrite morphological evolution from faceted tip to faceted tip after discontinuous change of anisotropy parameter $\epsilon_d$ at reduced time $t=400$. a) Decrease of $\epsilon_d$  from $\epsilon_d^{(1)}=0.1$ (red curve, $t=400$, faceted dendrite with $R=2.9$) to $\epsilon_d^{(2)}=0.05$  (blue curves from $t=400$ to $t=1000$, faceted dendrite). b) Increase of $\epsilon_d$  from $\epsilon_d^{(1)}=0.05$ (red curve, $t=400$, faceted dendrite) to $\epsilon_d^{(2)}=0.1$  (blue curves from $t=400$ to $t=1000$, faceted dendrite).  a-b) Insert: initial variation of radius $r(\theta)$ between $t=400$ and $t=402.5$.}
\label{fig:facet} 
\end{figure} 	

In order to assess the impact of capillary anisotropy shift on the morphological stability of the growing dendrite, the following procedure was implemented. The dendrite was first grown from a  small germ, using a first value $\epsilon_d^{(1)}$ of the capillary anisotropy parameter, until $t=400$. This duration was sufficient to reach the steady state regime.  Then, the capillary anisotropy parameter $\epsilon_d^{(1)}$ was updated to a second value $\epsilon_d^{(2)}$, while all other parameters were kept constant, as well as the temperature field, including at the interface ($T_i^{(1)}=T_i^{(2)}$). Three representative cases were addressed in this work:


\paragraph {Case 1:} transition from a smooth parabolic dendrite tip with underlying capillary anisotropy parameter $\epsilon_d^{(1)}<\epsilon_m$, to a second smooth dendrite tip $\epsilon_d^{(2)}<\epsilon_m$. 
The different stages of the process are displayed in figure \ref{fig:smooth}. Two subcases are prospected. $\epsilon_d$ is decremented from 0.02 to 0.01 on the one hand (see figure \ref{fig:C}), and incremented from 0.01 to 0.02 on the other hand (see figure \ref{fig:D}).

\paragraph {Case 2:} transition from a faceted dendrite tip with underlying capillary anisotropy parameter $\epsilon_d^{(1)}>\epsilon_m$, to a smooth dendrite tip $\epsilon_d^{(2)}<\epsilon_m$ (see figure \ref{fig:A}), and the opposite situation wherein $\epsilon_d$ is raised from 0.01 to 0.1  (see figure \ref{fig:B}).

\paragraph {Case 3:} transition from a faceted dendrite tip with underlying capillary anisotropy parameter $\epsilon_d^{(1)}>\epsilon_m$, to a second faceted dendrite tip $\epsilon_d^{(2)}>\epsilon_m$ (see figure \ref{fig:facet}). $\epsilon_d$ is alternately decreased from 0.1 to 0.05 (figure \ref{fig:E}) and increased from 0.05 to 0.1 (figure \ref{fig:F}).

In figures \ref{fig:smooth}, \ref{fig:facetsmooth} and \ref{fig:facet}, only the last step ($t=400$) of the first configuration dendrite contour ($\epsilon_d=\epsilon_d^{(1)}$)  is depicted (red curve). Herein, different growth stages from $t=400$ to $t=1000$ of the dendrite after shift of capillary anisotropy parameter to $\epsilon_d^{(2)}$ are shown (blue curves). On these curves, the red and green dots indicate the tip of the secondary branch and the center of the secondary hollow respectively. These dots are used as a guide for the eye, to trace back the initial variation of the initial state of configuration 2 from the last step of configuration 1 ($t=400$, red curve). The underlying idea is to connect the early morphological dynamics of the dendrite after capillary anisotropy disruption, to the outcoming ISB process, in order to assess the ISB mechanism summarized in equation \ref{C} and figure \ref{fig:theo}.

For that purpose, the early variations  $\delta r$ of the radius $r$ as defined in figure \ref{fig:tip} between dendrite contours at $t=400$ and $t=402.5$,  is plotted as a function of the normal angle $\theta$ in the insert on the right. This curve serves as a reliable indicator of the variation of the normal interface velocity $\delta v_n(\theta)$ resulting from capillary effects disruption, after integration on an infinitesimal time.


The value of the tip radius $R$ in the parabolic dendrite tip assumption is chosen case-wise. When the dendrite tip of configuration 1 is smooth ($\epsilon_d^{(1)}<\epsilon_m$), it is reasonably chosen the curvature radius of steady state configuration 1: $R=R^{(1)}$ (figures \ref{fig:C}, \ref{fig:D} and \ref{fig:B}). When the dendrite tip of configuration 1 is faceted, but the capillary anisotropy coefficient of configuration 2 leads to a smooth parabolic dendrite tip, the curvature of the freshly smoothed faceted tip at $t=402.5$ is used  (figure \ref{fig:A}). Finally, when both the dendrite tips of configurations 1 and 2 are faceted (figures \ref{fig:E} and \ref{fig:F}), we used the effective curvature radius $R_{\text{moy}}=10$.

In addition, the initial  variation of the position of the secondary branch tip and the secondary hollow, which is framed on the initial dendrite contour (red curve), is indicated in the insert. It was obtained by the local average of the early variation of radius $\delta r$ (red curve in the insert). Finally, the theoretical value of angles $\theta_1$ and  $\theta_2$ (and $\theta_0$ when relevant) are carried over the abscissa axis.


On the one hand, the decrease of $\epsilon_d$ induces the formation of one side branch at a normal angle $\theta_1$ spotted by a red dot on the $\theta$-axis in cases 1 and 2 (see insert in figures \ref{fig:C} and \label{fig:A}). This direction corresponds to the local maximum of the $(\theta,\delta r)$ curve in the same insert. The value of $\theta_1$ is close to the theoretical local maximum $\theta_1=0.48$ for Case 1, and $\theta_1=0.51$ for Case 2. However, the initial bump at $\theta\simeq\theta_1$ fails to produce side branching in case 3 (see \ref{fig:E}). On explanation might be the presence of two closely located deep minima in $\theta_0$ and $\theta_2$ in figure \ref{fig:1facet} (red curve), which could preclude the development of the initial interface variation in $\theta_1$.

On the other hand, the increment of $\epsilon_d$ in figures \ref{fig:D}, \label{fig:B} and \ref{fig:F} entails the curbing of the interface for normal directions close to a maximal curbing normal angle $\theta_1$. This direction is indicated by a green dot in the insert, pertaining to the local minimum of the $(\theta,\delta r)$ curve. Also, the increase of $\epsilon_d$ additionally begets the formation of a side-branch bump at a second  angle $\theta_2>\theta_1$ corresponding to the local maximum of the $(\theta,\delta r)$ curve  (red dot in the insert). This direction $\theta_2$ is close to its theoretical counterpart  $\theta_2=0.94$ for smooth dendrites (figure \ref{fig:D}). When faceting is involved (figures \ref{fig:B} and \ref{fig:F}), a significantly smaller value is obtained  $\theta_2\simeq 0.75$ drawing near to its theoretical counterpart  $\theta_2=0.65$. The smaller value of $\theta_2$ in case 2 and 3 compare to case 1 ($\theta_2=0.65$ in theory, and $\theta_2\simeq 0.75$ in cases 2 and 3 vs. $\theta_2=0.94$ in case 1)  is indeed likely to foster the enhanced sprouting of side-branches, insofar as the initial side-branch variation lies closer to the tip in  this configuration.

Noteworthy, the local minima $\theta_0$ and $\theta_2$ of $\delta v_n$ in figure \ref{fig:1facetsmooth} does not feed through the early deformation of the dendrite shape whenever $\epsilon_d$ is shrunk in  case 2. One explanation might be the negligible amplitude of the corresponding two minima in figure \ref{fig:smooth} (green curve) as compared to the maximum $\theta_1$.

One remarkable observation is that the location ($\theta_1$ or $\theta_2$) of the initial variation inducing side branching is to some extent independent of the initial morphology for the dendrite. This is particularly striking when the dendrite tip is faceted. Contrary to what was surmised in \cite{demange2017phase}, the ISB is not nucleated at the tip front facet corner, but at the fixed direction $\theta_{1,2}$. As for the initial bump at the tip front facet corner, it flattens and vanishes (see figure \ref{fig:A}).  

\section{Discussion}
\label{discussion}

The side branching observed in this work differs from noise ISB, wherein thermal fluctuations at the tip are converted in a train of periodic interface fluctuations whose period is set by the Mullins-Sekerka spectrum. These fluctuations are amplified as they propagate backward from the tip, until side branches are formed. Side branches hereafter compete, until the surviving ones reach the coarsening regime. In the end, side branches seldom reach the free growth regime. This characteristic behavior is envisioned in most phase-field studies equipped with noise induced side branching  \cite{kobayashi1993modeling,karma1999phase}. Things obviously go very differently in case of induced side branching. Only one side branch couple is formed, so that no competition between side branches occur, and the free growth regime can be reached. Besides, depending on the growth conditions change applied, side-branches can sprout precociously in the vicinity of the tip, thence exacerbating the subsequent development of the freely growing secondary branches. Albeit we did not check this point, one natural consequence of prematurely free growing side branches should be a growth law deviating from \cite{brener1995noise}.  


A second observation relates to the influence  of the choice of regularization algorithm  for unstable orientations on presently envisioned ISB in case of faceted growth. Indeed, alternative regularization procedures can be found in the literature for both capillary and kinetic faceting (see for instance \cite{bollada2018faceted} for a short review), including cusp removal \cite{debierre2003phase}, and the addition of higher derivative order terms to the surface energy function \cite{wheeler2006phase}. In this regards, one can thus argue that the present conclusions on ISB for faceted growth may reflect the choice of the selected regularization procedure, more than the underlying physics of the process. On this point, we suggest that numerical observations (figures \ref{fig:facetsmooth} and \ref{fig:facet}) and conclusions might not be radically altered, as we believe that it is the transition from stable to unstable crystal orientations at $|\theta|=\theta_m$ where the stiffness vanishes, that is responsible for the strong departure of the interface equilibrium temperature between configuration 1 and 2, which itself triggers side-branching. In particular, choosing to remove cusps in place of convexifying the inverse polar plot will only result in a slight variation of $\theta_{0,1,2}$. However, analytical curves displayed in figures \ref{fig:1facetsmooth} and \ref{fig:1facet} rely upon the parabolic tip approximation for the interface curvature. This simplification might be responsible for the discrepancy between expected values of  $\theta_1$ and  $\theta_2$ and the numerical values spotted in the insert of \ref{fig:facetsmooth} and \ref{fig:facet}. Moreover, this simplification might become even more pregnant with implications whenever the curvature should be added to the expression of the surface energy  \cite{di1992regularized,golovin1999modeling,golovin2001convective}. This question should be addressed in another study. More generally, the present work should also be extended to kinetic anisotropy in a first time, and then to a general variation of the dendrite morphology in a second time.



With all due caution, namely the difference of P\'eclet number and growth regime (diffusion limited growth vs. slow growth), origin of growth anisotropy (capillary vs. kinetic), as well as the peculiarities of smooth and rough faceting \cite{akutsu2021faceted}, the presently described mechanism for ISB might provide an alternative scenario for the sprouting of side branches in snowflake as was experimentally observed upon changing growth conditions in \cite{libbrecht2019snow}. In his book, Libbrecht ascribes ISB to a purely morphological origin. It is surmised that upon switching from a smooth to a faceted and then again to a smooth dendrite by means of supersaturation tuning would cause side-branches  to emerge at the freshly formed tip, in virtue of Berg's effect \cite{berg1938crystal}. Now, the present work suggests that under such disturbances, induced side-branches should form closer to the tip, and only in a second time migrate backwards from the tip. In addition, secondary branching might be achieved by alternative growth condition shifts, including certain transitions from smooth to smooth, faceted to faceted, and smooth to faceted dendrite tip.

\section{Conclusion}

In this work, the process of induced side branching caused by a long range disturbance of the growth conditions was prospected. An alternative mechanism for the secondary branching instability was proposed. It is suggested that side-branching can be the final outcome of a global deviation of the dendrite shape from its steady state morphology. Therefrom, the destabilization of the tip reflects the trade-off between non monotonic variations of interface temperature, surface energy, kinetic anisotropy and interface velocity to comply with the Gibbs Thomson equation. In this regards, the toy model of side branching induced by the tuning of capillary anisotropy effects was used for both smooth and faceted dendrite growth, as it can be easily addressed analytically and numerically through faceted phase-field simulations. In this manner, it was envisioned that the induced secondary branching instability could ensue a host of capillary anisotropy effects changes, with specific signatures. The present study should now be extended to more realistic growth condition changes, and meshed within the more general framework of bifurcation in a side-branch surface \cite{martin1987origin,galenko1997bifurcations}.

\enlargethispage{20pt}





\bibliographystyle{unsrt}

\begin{thebibliography}{49}

\bibitem{arivanandhan2019crystallization}
Mukannan Arivanandhan, Genki Takakura, D~Sidharth, Maeda Kensaku, Keiji Shiga,
  Haruhiko Morito, and Kozo Fujiwara.
\newblock Crystallization and re-melting of si1-xgex alloy semiconductor during
  rapid cooling.
\newblock {\em Journal of Alloys and Compounds}, 798:493--499, 2019.

\bibitem{libbrecht2005physics}
K.~G Libbrecht.
\newblock The physics of snow crystals.
\newblock {\em Rep. Prog. Phys.}, 68(4):855, 2005.

\bibitem{shimada2018rapid}
Wataru Shimada and Shohei Furukawa.
\newblock Rapid growth of ice crystal dendrite tips in dilute solution of
  trehalose.
\newblock {\em Journal of Crystal Growth}, 493:25--29, 2018.

\bibitem{PEPI}
DV~Alexandrov and AP~Malygin.
\newblock Coupled convective and morphological instability of the inner core
  boundary of the earth.
\newblock {\em Physics of the Earth and Planetary Interiors},
  189(3-4):134--141, 2011.

\bibitem{AG_ICB2013}
DV~Alexandrov and PK~Galenko.
\newblock Selection criterion for the growing dendritic tip at the inner core
  boundary.
\newblock {\em Journal of Physics A: Mathematical and Theoretical},
  46(19):195101, 2013.

\bibitem{hoyt2003}
JJ~Hoyt, Mark Asta, and Alain Karma.
\newblock Atomistic and continuum modeling of dendritic solidification.
\newblock {\em Materials Science and Engineering: R: Reports}, 41(6):121--163,
  2003.

\bibitem{li1999evolution}
Q~Li and C~Beckermann.
\newblock Evolution of the sidebranch structure in free dendritic growth.
\newblock {\em Acta materialia}, 47(8):2345--2356, 1999.

\bibitem{hansen2002dendritic}
George Hansen, Shan Liu, Shu-Zu Lu, and Angus Hellawell.
\newblock Dendritic array growth in the systems nh4cl--h2o and [ch2cn] 2--h2o:
  steady state measurements and analysis.
\newblock {\em Journal of crystal growth}, 234(4):731--739, 2002.

\bibitem{umantsev2011origin}
Alexander Umantsev.
\newblock On the origin of biological functions.
\newblock {\em Journal of Modern Physics}, 2(6):602--614, 2011.

\bibitem{ivantsov1952growth}
GP~Ivantsov.
\newblock On a growth of spherical and needle-like crystals of a binary alloy.
\newblock In {\em Dokl. Akad. Nauk SSSR}, volume~83, pages 573--575, 1952.

\bibitem{kessler1987growth}
David~A Kessler and Herbert Levine.
\newblock Growth velocity of three-dimensional dendritic crystals.
\newblock {\em Physical Review A}, 36(8):4123, 1987.

\bibitem{barbieri1989predictions}
A~Barbieri and JS~Langer.
\newblock Predictions of dendritic growth rates in the linearized solvability
  theory.
\newblock {\em Physical Review A}, 39(10):5314, 1989.

\bibitem{Brener1990}
EA~Brener.
\newblock Effects of surface energy and kinetics on the growth of needle-like
  dendrites.
\newblock {\em Journal of crystal growth}, 99(1-4):165--170, 1990.

\bibitem{brener1991pattern}
EA~Brener and VI~Mel'Nikov.
\newblock Pattern selection in two-dimensional dendritic growth.
\newblock {\em Advances in Physics}, 40(1):53--97, 1991.

\bibitem{alexandrov2016selection}
DV~Alexandrov, DA~Danilov, and PK~Galenko.
\newblock Selection criterion of a stable dendrite growth in rapid
  solidification.
\newblock {\em International Journal of Heat and Mass Transfer}, 101:789--799,
  2016.

\bibitem{alexandrov2013selection}
Dmitri~V Alexandrov and Peter~K Galenko.
\newblock Selection criterion of stable dendritic growth at arbitrary
  p{\'e}clet numbers with convection.
\newblock {\em Physical Review E}, 87(6):062403, 2013.

\bibitem{alexandrov2017dendritic}
DV~Alexandrov and PK~Galenko.
\newblock Dendritic growth with the six-fold symmetry: theoretical predictions
  and experimental verification.
\newblock {\em Journal of Physics and Chemistry of Solids}, 108:98--103, 2017.

\bibitem{kao2020stable}
A~Kao, LV~Toropova, I~Krastins, G~Demange, DV~Alexandrov, and PK~Galenko.
\newblock A stable dendritic growth with forced convection: A test of theory
  using enthalpy-based modeling methods.
\newblock {\em JOM}, 72(9):3123--3131, 2020.

\bibitem{horvay1961dendritic}
G~Horvay and JW~Cahn.
\newblock Dendritic and spheroidal growth.
\newblock {\em Acta Metallurgica}, 9(7):695--705, 1961.

\bibitem{toropova2020non}
LV~Toropova, PK~Galenko, DV~Alexandrov, M~Rettenmayr, A~Kao, and G~Demange.
\newblock Non-axisymmetric growth of dendrite with arbitrary symmetry in two
  and three dimensions: sharp interface model vs phase-field model.
\newblock {\em The European Physical Journal Special Topics},
  229(19):2899--2909, 2020.

\bibitem{JCG2021}
LV~Toropova, E~Titova, DV~Alexandrov, PK~Galenko, A~Kao, and G~Demange.
\newblock Dendritic growth of ice crystals: a test of theory with experiments.
\newblock {\em Journal of Crystal Growth}, 2021.

\bibitem{PTA2021}
DV~Alexandrov, LV~Toropova, E~Titova, A~Kao, G~Demange, PK~Galenko, and
  M~Rettenmayr.
\newblock The shape of dendritic tips: a test of theory with computations and
  experiments.
\newblock {\em Philosophical Transactions of the Royal Society of London.
  Series A: Mathematical, Physical and Engineering Sciences}, 2021.

\bibitem{pieters1986noise}
Roger Pieters and JS~Langer.
\newblock Noise-driven sidebranching in the boundary-layer model of dendritic
  solidification.
\newblock {\em Physical review letters}, 56(18):1948, 1986.

\bibitem{langer1987dendritic}
JS~Langer.
\newblock Dendritic sidebranching in the three-dimensional symmetric model in
  the presence of noise.
\newblock {\em Physical Review A}, 36(7):3350, 1987.

\bibitem{barber1987dynamics}
Michael~N Barber, Angelo Barbieri, and JS~Langer.
\newblock Dynamics of dendritic sidebranching in the two-dimensional symmetric
  model of solidification.
\newblock {\em Physical Review A}, 36(7):3340, 1987.

\bibitem{brener1995noise}
Efim Brener and Dmitri Temkin.
\newblock Noise-induced sidebranching in the three-dimensional nonaxisymmetric
  dendritic growth.
\newblock {\em Physical Review E}, 51(1):351, 1995.

\bibitem{glicksman2007deterministic}
Martin~E Glicksman, John~S Lowengrub, Shuwang Li, and Xiangrong Li.
\newblock A deterministic mechanism for dendritic solidification kinetics.
\newblock {\em JOM}, 59(8):27, 2007.

\bibitem{he2003heat}
X~He, PW~Fuerschbach, and T~DebRoy.
\newblock Heat transfer and fluid flow during laser spot welding of 304
  stainless steel.
\newblock {\em Journal of Physics D: Applied Physics}, 36(12):1388, 2003.

\bibitem{xie2014growth}
Yu~Xie, Hongbiao Dong, and Jonathan Dantzig.
\newblock Growth of secondary dendrite arms of fe--c alloy during transient
  directional solidification by phase-field method.
\newblock {\em ISIJ international}, 54(2):430--436, 2014.

\bibitem{gonzalez2004deterministic}
Ricard Gonzalez-Cinca, Yves Couder, J~Maurer, and Aurora Hernandez-Machado.
\newblock Deterministic versus noisy behavior in sidebranching.
\newblock In {\em Noise in Complex Systems and Stochastic Dynamics II}, volume
  5471, pages 280--288. International Society for Optics and Photonics, 2004.

\bibitem{libbrecht2019snow}
Kenneth~G. Libbrecht.
\newblock Snow crystals, 2019.

\bibitem{alexandrov2018thermo}
Dmitri~V Alexandrov, Peter~K Galenko, and Lyubov~V Toropova.
\newblock Thermo-solutal and kinetic modes of stable dendritic growth with
  different symmetries of crystalline anisotropy in the presence of convection.
\newblock {\em Philosophical Transactions of the Royal Society A: Mathematical,
  Physical and Engineering Sciences}, 376(2113):20170215, 2018.

\bibitem{mcfadden1993phase}
G.~B. McFadden, A.~A. Wheeler, R.~J. Braun, S.~R. Coriell, and R.~F. Sekerka.
\newblock Phase-field models for anisotropic interfaces.
\newblock {\em Phys. Rev. E}, 48(3):2016, 1993.

\bibitem{eggleston2001phase}
Joshua~J Eggleston, Geoffrey~B McFadden, and Peter~W Voorhees.
\newblock A phase-field model for highly anisotropic interfacial energy.
\newblock {\em Physica D: Nonlinear Phenomena}, 150(1-2):91--103, 2001.

\bibitem{demange2017phase}
Gilles Demange, Helena Zapolsky, Renaud Patte, and Marc Brunel.
\newblock A phase field model for snow crystal growth in three dimensions.
\newblock {\em npj Comput. Mater.}, 3:1, 2017.

\bibitem{demange2017growth}
G~Demange, H~Zapolsky, R~Patte, and M~Brunel.
\newblock Growth kinetics and morphology of snowflakes in supersaturated
  atmosphere using a three-dimensional phase-field model.
\newblock {\em Physical Review E}, 96(2):022803, 2017.

\bibitem{ramirez2004phase}
JC~Ramirez, C~Beckermann, As~Karma, and H-J Diepers.
\newblock Phase-field modeling of binary alloy solidification with coupled heat
  and solute diffusion.
\newblock {\em Physical Review E}, 69(5):051607, 2004.

\bibitem{kobayashi1993modeling}
Ryo Kobayashi.
\newblock Modeling and numerical simulations of dendritic crystal growth.
\newblock {\em Physica D: Nonlinear Phenomena}, 63(3-4):410--423, 1993.

\bibitem{karma1999phase}
Alain Karma and Wouter-Jan Rappel.
\newblock Phase-field model of dendritic sidebranching with thermal noise.
\newblock {\em Physical review E}, 60(4):3614, 1999.

\bibitem{bollada2018faceted}
PC~Bollada, PK~Jimack, and AM~Mullis.
\newblock Faceted and dendritic morphology change in alloy solidification.
\newblock {\em Computational Materials Science}, 144:76--84, 2018.

\bibitem{debierre2003phase}
Jean-Marc Debierre, Alain Karma, Franck Celestini, and Rahma Gu{\'e}rin.
\newblock Phase-field approach for faceted solidification.
\newblock {\em Physical Review E}, 68(4):041604, 2003.

\bibitem{wheeler2006phase}
AA~Wheeler.
\newblock Phase-field theory of edges in an anisotropic crystal.
\newblock {\em Proceedings of the Royal Society A: Mathematical, Physical and
  Engineering Sciences}, 462(2075):3363--3384, 2006.

\bibitem{di1992regularized}
Antonio Di~Carlo, Morton~E Gurtin, and Paolo Podio-Guidugli.
\newblock A regularized equation for anisotropic motion-by-curvature.
\newblock {\em SIAM Journal on Applied Mathematics}, 52(4):1111--1119, 1992.

\bibitem{golovin1999modeling}
AA~Golovin, SH~Davis, and AA~Nepomnyashchy.
\newblock Modeling the formation of facets and corners using a convective
  cahn--hilliard model.
\newblock {\em Journal of crystal growth}, 198:1245--1250, 1999.

\bibitem{golovin2001convective}
AA~Golovin, AA~Nepomnyashchy, Stephen~H Davis, and MA~Zaks.
\newblock Convective cahn-hilliard models: From coarsening to roughening.
\newblock {\em Physical Review Letters}, 86(8):1550, 2001.

\bibitem{akutsu2021faceted}
Noriko Akutsu.
\newblock Faceted-rough surface with disassembling of macrosteps in
  nucleation-limited crystal growth.
\newblock {\em Scientific Reports}, 11(1):1--11, 2021.

\bibitem{berg1938crystal}
W.~F. Berg.
\newblock Crystal growth from solutions.
\newblock In {\em Proc. R. Soc. A}, volume 164, pages 79--95, 1938.

\bibitem{martin1987origin}
Olivier Martin and Nigel Goldenfeld.
\newblock Origin of sidebranching in dendritic growth.
\newblock {\em Physical Review A}, 35(3):1382, 1987.

\bibitem{galenko1997bifurcations}
PK~Galenko, MD~Krivilyov, and SV~Buzilov.
\newblock Bifurcations in a sidebranch surface of a free-growing dendrite.
\newblock {\em Physical Review E}, 55(1):611, 1997.

\end{thebibliography}

\end{document}